\documentclass[reprint,pra,amsmath,amssymb, aps,longbibliography]{revtex4-1}

\usepackage{comment}
\usepackage{xcolor}
\usepackage{graphicx}
\usepackage{dcolumn}
\usepackage{bm}
\usepackage[colorlinks,linkcolor=red,citecolor=blue,urlcolor=red]{hyperref}

\renewcommand{\t}[1]{\mathrm{#1}}

\AtBeginDocument{%
	\newwrite\bibnotes
	\def\bibnotesext{Notes.bib}
	\immediate\openout\bibnotes=\jobname\bibnotesext
	\immediate\write\bibnotes{@CONTROL{REVTEX41Control}}
	\immediate\write\bibnotes{@CONTROL{%
			apsrev41Control,author="08",editor="1",pages="1",title="0",year="1"}}
	\if@filesw
	\immediate\write\@auxout{\string\citation{apsrev41Control}}%
	\fi
}

\newcommand\numberthis{\addtocounter{equation}{1}\tag{\theequation}}

\begin{document}
	
	
	\title{\textcolor{black}{High-cooperativity torsional optomechanics with an optical lever}}
	\title{\textcolor{black}{Quantum-limited optical lever measurements of a torsion oscillator}}
	
	\author{C. M. Pluchar, A. R. Agrawal, and D. J. Wilson}
	
	\affiliation{Wyant College of Optical Sciences, University of Arizona, Tucson, AZ 85721, USA}
	
	\date{\today}
	
	\begin{abstract}
		The optical lever is a precision displacement sensor with broad applications. 
		In principle, it can track the motion of a mechanical oscillator with added noise at the Standard Quantum Limit (SQL); however, demonstrating this performance requires an oscillator with an exceptionally high torque sensitivity, or, equivalently, zero-point angular displacement spectral density.  Here, we describe optical lever measurements on Si$_3$N$_4$ nanoribbons possessing $Q>3\times 10^7$ torsion modes with torque sensitivities of $10^{-20}\,\text{N m}/\sqrt{\text{Hz}}$ and zero-point displacement spectral densities of $10^{-10}\,\text{rad}/\sqrt{\text{Hz}}$.
		Compensating aberrations and leveraging immunity to classical intensity noise, we realize angular displacement measurements with imprecisions 20 dB below the SQL and demonstrate feedback cooling, using a position-modulated laser beam as a torque actuator, from room temperature \mbox{to $\sim5000$} phonons.  Our study signals the potential for a new class of  torsional quantum optomechanics.

	\end{abstract}
	
	\maketitle
	
	Optical metrology enables precise tracking of mechanical oscillators. This is a key paradigm in the search for new physics, as mechanical oscillators can transduce weak forces such as radiation pressure \cite{nichols1903pressure}, gravitational waves \cite{abbott2016observation}, and the electrostatic force \cite{coulombFirstDissertationElectricity1788} into tangible displacements. 
	In the last decade, optomechanical measurements have reached a regime where their added noise is limited by quantum fluctuations of the light field, including radiation pressure shot noise \cite{purdy2013observation}. 
	Subsequently, squeezed light \cite{aasi2013enhanced} and backaction evading
	\cite{shomroni2019optical, ganapathy2023broadband} techniques have provided reductions in quantum noise, leading to force and displacement measurements below the Standard Quantum Limit (SQL) \cite{mason2019continuous, jia2024ligo}.
	This has allowed for a new generation of fundamental physics experiments with even greater sensitivity, which may be useful for dark matter searches \cite{carney2021mechanical} and novel tests of gravity \cite{rademacher2020quantum}.

	While theory and experiment in optical displacement measurement has focused on interferometry, the quantum limits of alternative techniques, such as the optical lever (OL), have been largely ignored.  The OL is notable because of its long history as a precision measurement tool~\cite{jones1961some}, including direct measurements of radiation pressure~\cite{nichols1903pressure} and gravity~\cite{adelberger2009torsion, westphal2021measurement}, and its employment in commercial atomic force microscopes \cite{alexander1989atomic}. There is moreover no fundamental advantage to interferometry, as previous analyses indicate that the displacement sensitivity of the OL is on par~\cite{putman1992detailed, putman1992theoretical}. 
	While the quantum limit of the OL and closely related lateral beam displacement problem has been studied~\cite{barnett2003ultimate, delaubert2006quantum, hsu2004optimal, he2024optimum}, 
	including enhancement using squeezed light \cite{treps2003quantum,delaubert2006quantum,pooser2015ultrasensitive}, radiation pressure quantum backaction was not accounted for. As contributions of both imprecision and backaction enforce the SQL, this remains an unexplored regime.  A notable exception is the recent demonstration of classical backaction evasion in an OL~\cite{hao2024back}, which has heavily influenced our study.
	
	\color{black}
	\begin{figure}[t]
		\centering
		\includegraphics[width=0.85\columnwidth]{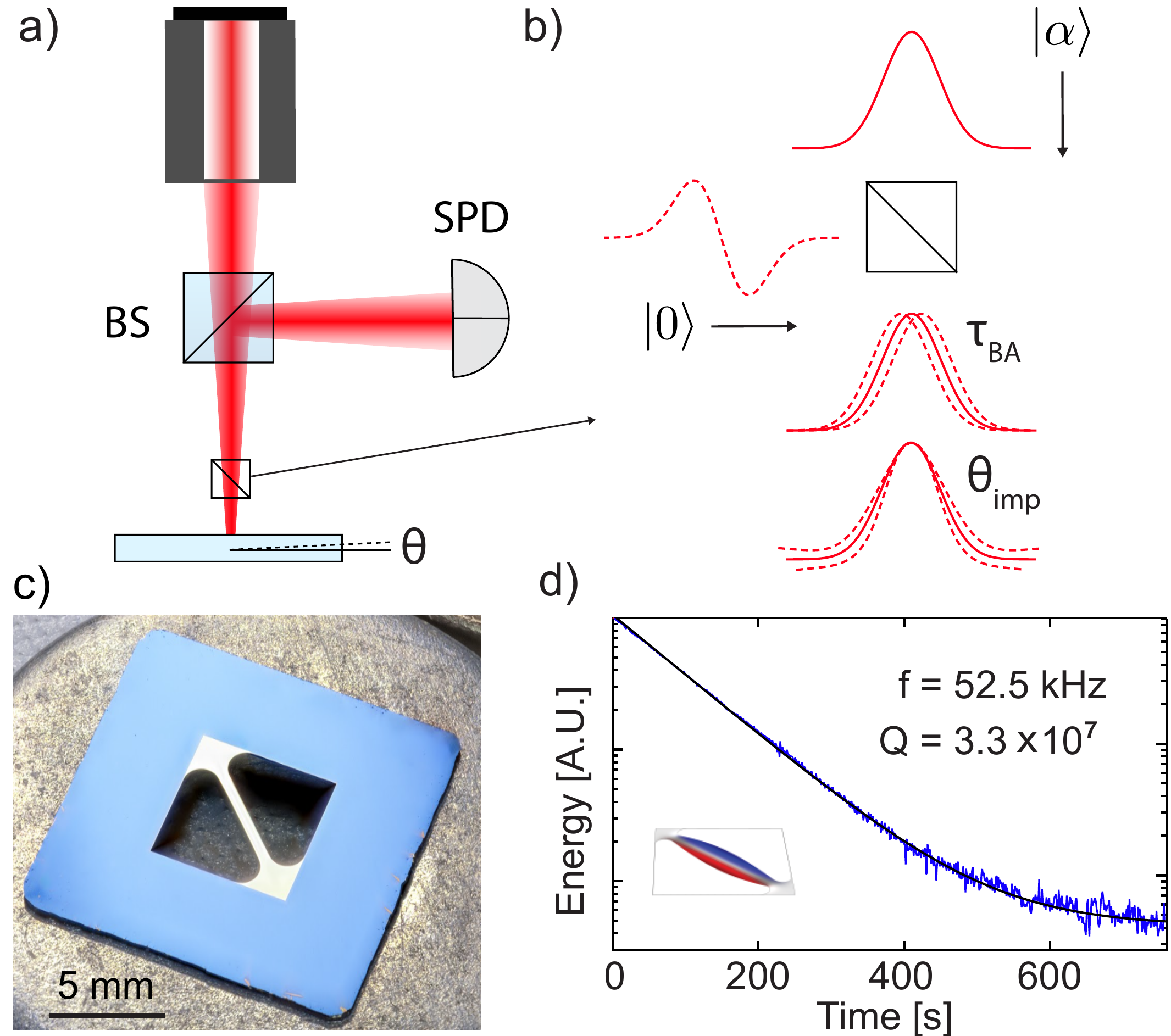}
		\caption{(a) Sketch of an optical lever measuring angular displacement $\theta$ using a split photodetector (SPD). b) Quantum noise model: the incident field in an HG$_{00}$ coherent state beats against the HG$_{10}$ vacuum, resulting in uncertainty in the incidence angle $\theta$ (imprecision) and position $x$ (backaction) due to phase and amplitude vacuum noise, respectively.  (c) Photo of a 400-$\mu$m-wide Si$_3$N$_4$ nanoribbon \cite{pratt2023nanoscale}  (d) Ringdown of the nanoribbon's fundamental torsion mode (inset).}
		\label{fig:OL}
		\vspace{-3mm}
	\end{figure}
	
	In this Letter, we present a platform to explore the quantum limits of the OL and angular displacement measurements. 
	Two challenges need addressing: first, the thermal torque noise of the mechanical oscillator $S_\tau^\t{th}$ (here expressed as a single-sided power spectral density evaluated at resonance) needs to be made comparable to the quantum backaction torque $S^\t{BA}_\tau$. Second, the optical receiver must possess a high quantum efficiency $\eta$, in the sense that the product of the backaction and measurement imprecision $S_\theta^\t{imp}$ approaches the Heisenberg limit $S_\theta^\t{imp} S^\t{BA}_\tau  = \hbar^2/\eta \ge\hbar^2$  or equivalently, the total measurement noise approaches the SQL, $S_\theta^\t{imp} + S^\t{BA}_\theta  \ge S_\theta^\t{ZP}$, where $S_\theta^\t{ZP}$ is the oscillator's zero-point motion \cite{clerkIntroductionQuantumNoise2010}.
	
	To address these challenges, we probe Si$_3$N$_4$ nanoribbons \cite{pratt2023nanoscale} possessing a $Q>3\times 10^7$ torsion modes  with thermal torques of $S_\tau^\t{th} \sim (10^{-20}\,\t{N m}/\sqrt{\t{Hz}})^2$ and zero-point spectral densities of $S_\theta^\t{ZP} \sim (10^{-10}\,\t{rad}/\sqrt{\t{Hz}})^2$. 
	For the receiver, we use a split photodetector, which is known to produce a near-ideal $\eta =2 / \pi$ OL measurement \cite{hsu2004optimal,fradgley2022quantum}.  We also carefully account for aberrations stemming from the finite size and curvature of the nanoribbon, and exploit access to large optical powers afforded by the relative immunity of the OL to classical intensity noise. 

	A sketch of the experiment is shown in Fig. \ref{fig:OL}a. Light from a laser (wavelength $\lambda$) in the fundamental Hermite-Gaussian (HG$_{00}$) mode is focused onto the ribbon to spot size ($1/e^2$ intensity radius) $w_0$, corresponding to diffraction angle $\theta_\t{D} = \lambda/(\pi w_0)$. In the small displacement limit $\theta \ll \theta_\t{D}$, the field reflected from the ribbon can be written as a superposition of HG$_{00}$ and HG$_{10}$ modes \cite{enomoto2016standard,hao2024back}
	\begin{equation}
		\label{eq:eRef}
		E_\t{r} \approx  A_{00} u_{00} + (A^\theta_{10} + A^\t{vac}_{10}) u_{10}
	\end{equation}
	where here $u_{00}$ ($u_{10}$) is the HG$_{00}$ (HG$_{10}$) modeshape and $A^\theta_{10} = 2i \theta/\theta_\t{D}$ is the amplitude of the HG$_{10}$ mode.

	A split photodetector (SPD) placed in the far field of the ribbon acts as a $\t{HG}_\t{10}$ mode sorter, producing a photocurrent proportional to $A_\t{10}^\theta$ and therefore $\theta$.  
	In Eq. \ref{eq:eRef}, we have included a term $A^\t{vac}_{10}$ representing vacuum fluctuations of the HG$_{10}$ mode.  
	As illustrated in Fig. \ref{fig:OL}b, these fluctuations produce angular and lateral beam displacement noise, yielding imprecision $S_\theta^\t{imp}$ and backaction $S_\tau^\t{BA}$, respectively. Referred to an apparent angular displacement, the total SPD output can be written
	\begin{subequations}\begin{align}
			S_\theta[\omega] & =  S_\theta^\t{imp}+\left|\chi_\t{m}[\omega]\right|^2 \left(S_\tau^\t{BA}+ S_\tau^\t{th}\right)+S_\theta^\t{ZP}[\omega]\\
		\label{eq:eq2}	&= S_\theta^\t{imp}+S_\theta^\t{BA}[\omega] + S_\theta^\t{th}[\omega]+S_\theta^\t{ZP}[\omega]
	\end{align}\end{subequations}
	where $\chi_\t{m} [\omega] =  I^{-1} / (\omega^2 - \omega_\t{m}^2 - i \gamma_\t{m} \omega)$, $\omega_\t{m}, \gamma_\t{m},$ and $I$ are the mechanical susceptibility, frequency, damping rate, and moment of inertia of the torsion mode, respectively, $S_\tau^\t{th} \approx 4 k_\t{B} T I \gamma_\t{m}$ is the thermal torque in the high temperature limit ($T\gg\hbar\omega_\t{m}/k_B$); and $S_\theta^\t{BA}$ and $S_\theta^\t{th}$ are the backaction and thermal displacement, respectively.
	
	As shown in \cite{pratt2023nanoscale,hao2024back, pluchar2024imaging} and in the Appendix, placing the SPD in the far field yields an imprecision
	\begin{equation}
		\label{eq:shotNoiseImpOL}
		S_\theta^\t{imp} \approx \frac{\theta_\t{D}^2}{8N}\frac{\pi   }{2 \eta_\t{d}} 
		= \frac{1}{w_0^2}\frac{\hbar c \lambda}{ 4\pi P }\frac{\pi   }{2 \eta_\t{d}} 
	\end{equation}
	where $N$ ($P$) is the photon flux (optical power) on the photodetector and $\eta_\t{d}$ is its quantum efficiency.
	
	Likewise, radiation pressure backaction torque on the ribbon can be expressed as \cite{hao2024back, pluchar2024imaging} (see Appendix)
	\begin{equation}\label{eq:StauQBA}
		S_\tau^\t{BA} = \frac{8 N }{\theta_\t{D}^2}\frac{w^2}{w_0^2} \hbar^2  = w^2 \frac{4\pi\hbar P}{c \lambda } 
	\end{equation}
	where $w$ is the spot size on the ribbon. 
	
	\begin{figure}[t!]
		\includegraphics[width=1\columnwidth]{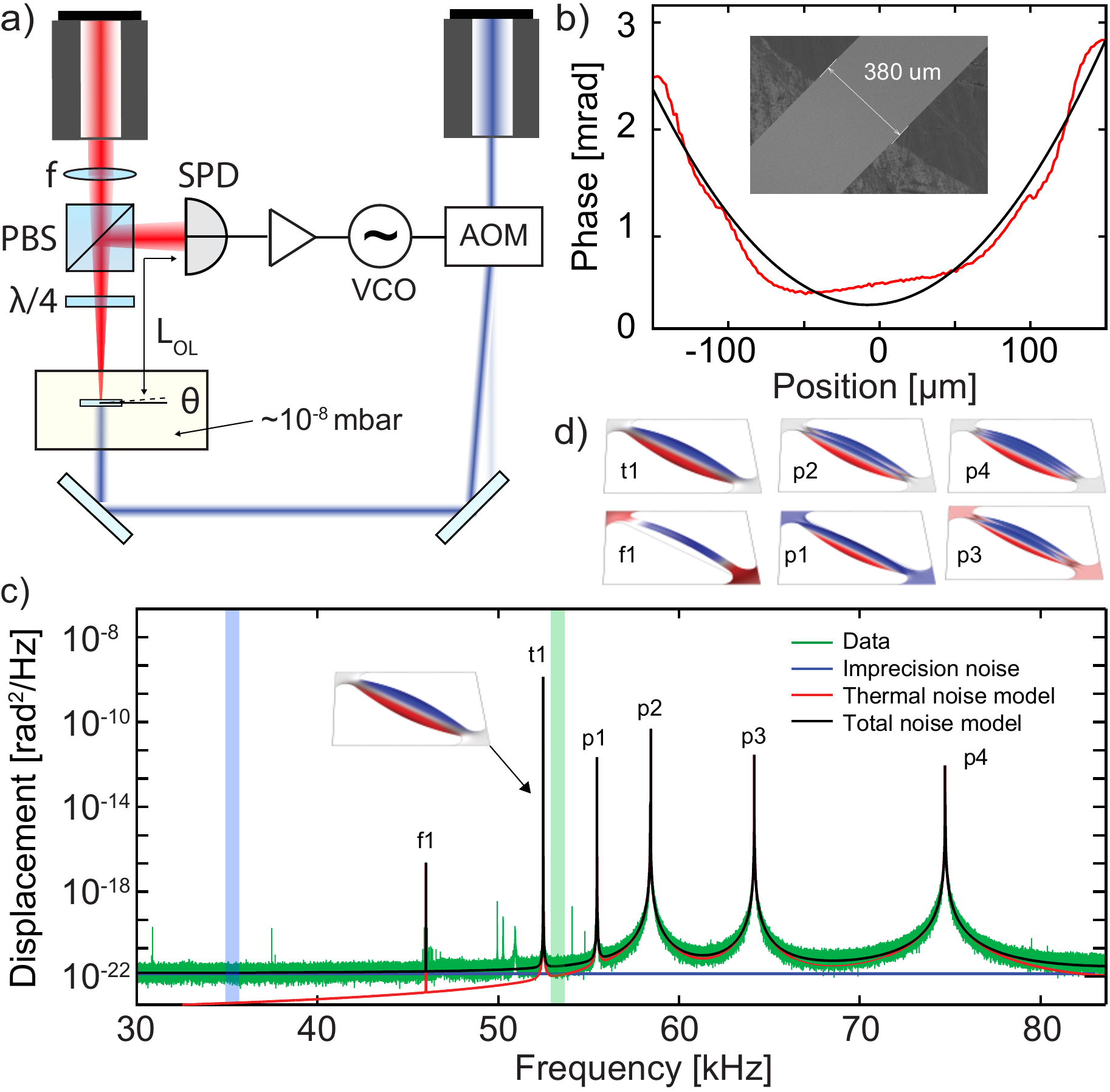}
		\caption{(a) Experimental setup. The nanoribbon is housed in a vacuum chamber at $10^{-8}$ mbar. An auxiliary acousto-optic modulated (AOM) beam is used for radiation pressure backaction simulation and feedback control. (b) White light profile of the ribbon near its midpoint, exhibiting effective parabolic curvature. (c) Typical displacement signal showing the torsion mode of interest (t1), ``potato chip'' modes (p1, p2, p3, p4), and a weakly coupled flexural mode (f1). The total noise model includes thermal noise and imprecision noise. (d) Finite element simulations of the various ribbon modes.}
		\label{fig:fig2}
		\vspace{-3mm}
	\end{figure}
	
	Combining Eqs. \ref{eq:shotNoiseImpOL} and \ref{eq:StauQBA}, the imprecision-backaction product for the optical lever can be written
	\begin{equation}
		S_{\theta}^{\t{imp}} S_{\tau}^{\t{BA}} =  \frac{\hbar^2}{\eta_\t{SD} \eta_\t{d}} \frac{w^2}{w_0^2},
	\end{equation}
	where $\eta_\t{SD} = 2/\pi$ characterizes the intrinsic nonideality of the SPD, stemming from its inability to distinguish $\t{HG}_{00}$ and $\t{HG}_{10}$ modes \cite{fradgley2022quantum,barnett2003ultimate, hsu2004optimal}. 
	The final term $w^2/w_0^2$ corresponds to excess backaction if the spot size on the ribbon is larger than the beam waist, and implies that focusing on the ribbon ($w = w_0$) gives the optimal imprecision \cite{putman1992detailed}, and imprecision-backaction product.
	
	\begin{figure*}[ht!]
		\includegraphics[width=2\columnwidth]{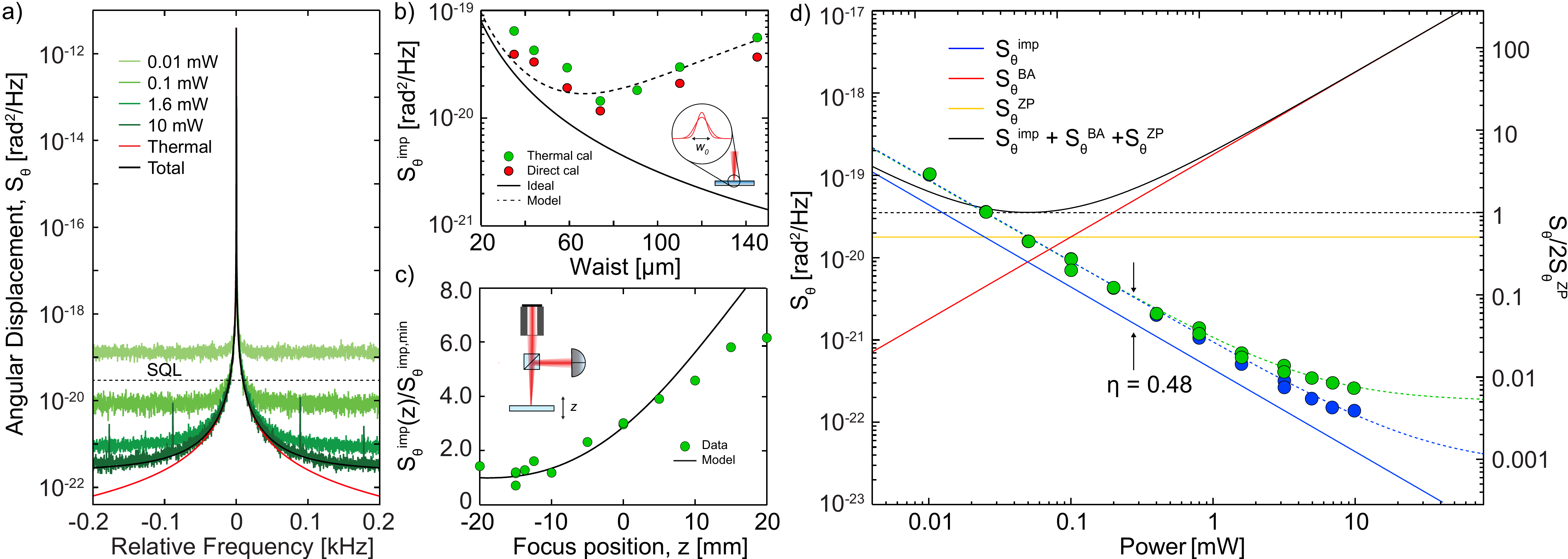}
		\caption{(a) Optical lever measurement of fundamental torsion mode for various  probe powers $P$.  Overlaid are models of thermal noise (red) and total noise for $P = 10$ mW (black). (b) Imprecision (noise floor) versus probe waist $w_0$ for $P = 100\;\mu\t{W}$. Green and red points are calibrated by bootstrapping to the thermal model (green) and measuring the response of the SD to a lateral displacement (red) \cite{pratt2023nanoscale}. (c) Imprecision versus beam focus position $z$ for $w_0 = 60\;\mu$m, normalized to the inferred minimum at $z = -0.2$ cm. (d) Imprecision versus $P$ for measurements as in (a).  Blue and green points are averages over far-off-resonant (blue) and near-resonant (green) frequency bands in Fig. \ref{fig:fig2}c. Overlaid are models of imprecision (blue), backaction (red), zero-point-motion (green), and their sum (black) for an ideal optical lever. 
			Green and blue dashed lines are fits consistent with a measurement efficiency $\eta = 48\%$ and an extraneous imprecision 22 and 30 dB below the SQL, respectively.}
		\label{fig:Fig3}
		\vspace{-3mm}
	\end{figure*}
	
	Details of the experiment are shown in Fig. \ref{fig:fig2}a. For the OL, we use $\lambda = 850$ nm light from a Ti-Sapphire laser (red).  The light is passed through an optical fiber followed by a collimating lens (not shown) to produce a near diffraction-limited HG$_{00}$ beam. A second lens ($f$) focuses the beam on the ribbon, and the return beam is directed to a SPD via a polarizing beamsplitter (PBS). The beam waist $w_0$, focal position relative to the sample $z$ and detector-sample separation (optical lever arm) $L_\t{OL}$ are important parameters for optimizing sensitivity.  Nominally, we arrange the setup so that $2w_0<w_\t{r}$, $z\lesssim z_0$, and $L_\t{OL}\gg z_0$, where $w_\t{r}$ is the sample (ribbon) width and $z_0 = \pi w_0^2/\lambda$ is the beam's Rayleigh length.

	In addition to the OL, we introduce an auxiliary position-modulated, 633 nm laser as a radiation pressure torque actuator.  Following \cite{hao2024back}, position-modulation is achieved by passing the beam through a frequency-modulated acousto-optic modulator.  
	A dichroic filter (not shown) is used to isolate the SPD from this laser.

	Our mechanical oscillator is a Si$_3$N$_4$ nanoribbon with length $L \approx 7$ mm, width $w_\t{r} = 400\;\mu$m, thickness $h = 75$ nm (Fig. \ref{fig:OL}c), fundamental torsion mode frequency $\omega_\t{m} = 2\pi\times 52.5\;\t{kHz}\approx (\pi/L)\sqrt{\sigma/\rho}$, and finite-element-simulated moment of inertia $I = 3.8\times 10^{-18}\;\t{kg}\;\t{m^2}\approx \rho  L hw_\t{r}^3/24$, where $\sigma\approx 0.85$ GHz and $\rho\approx 2700\;\t{kg}/\t{m}^3$ are the ribbon tensile stress and density, respectively. Previously \cite{pratt2023nanoscale}, we found that strain-induced dissipation dilution in these ribbons yield torsional $Q$ factors as high as $Q_0\sigma w_\t{r}^2/ (E h^2)\approx 10^8$, where $E$ and $Q_0$ are the ribbon elastic modulus and intrinsic $Q$. This is attractive because it implies access to exceptionally high torque sensitivities and zero-point spectral densities through the scaling laws $S_\tau^\t{th} = 4 k_B TI\omega_\t{m}/Q \propto h^3 w_\t{r} L/Q_0$ and $S_\theta^\t{ZP} = 2\hbar Q/(I\omega_\t{m}^2)\propto Q_0/(h^3 w_\t{r} L)$ \cite{pratt2023nanoscale}. Specifically, for the device used in this study, we measure $Q = 3.3\times 10^7$ via ringdown (Fig. \ref{fig:fig2}c), corresponding to $S_\theta^\t{th} \approx  (2.5\times 10^{-20}\;\t{N m}/\sqrt{\t{Hz}})^2$ and  $S_\theta^\t{ZP} = (1.3 \times 10^{-10}\;\t{rad}/\sqrt{\t{Hz}})^2$.
	
	Figure \ref{fig:Fig3} shows a set of experiments aimed at optimizing the efficiency of an OL measurement performed on the fundamental torsion mode of a nanoribbon.  First, we leverage the waist size dependence $S_\theta^\t{imp}\propto w_0^{-2}$ (Eq. \ref{eq:shotNoiseImpOL}) to reduce imprecision for a fixed power $P$. 
	Fig. \ref{fig:Fig3}b shows a compilation of $P = 100\;\mu\t{W}$ measurements with different waist sizes, by varying $f$. For $w_0 \le 50\,\mu\t{m}$, imprecision scales as $w_0^{-2}$ as expected.  For $w_0 \gtrsim 60\,\mu\t{m}$, it increases.  We attribute this discrepancy to two sources of extra diffraction: (1) the finite ribbon width results in clipping, and (2) the ribbon imparts a position-dependent phase shift due to the photoelastic effect. Fig. \ref{fig:fig2}d shows a white light interferogram of the ribbon cross-section, fit to a polynomial.
	The dominant fit parameter is quadratic, implying that the ribbon acts like a parabolic reflector with a radius of curvature $R_\t{r}\approx 3$ cm. 
	The orange curve in Fig. \ref{fig:Fig3}b is Fraunhofer diffraction model accounting for both effects (see Appendix). The model fits the data well, and implies that instead of decreasing monotonically with $w_0$,
	$S_\theta^\t{imp}$ is minimized for our device at $w_0 \approx 60\;\mu$m.

	\begin{figure*}  
		\includegraphics[width=2\columnwidth]{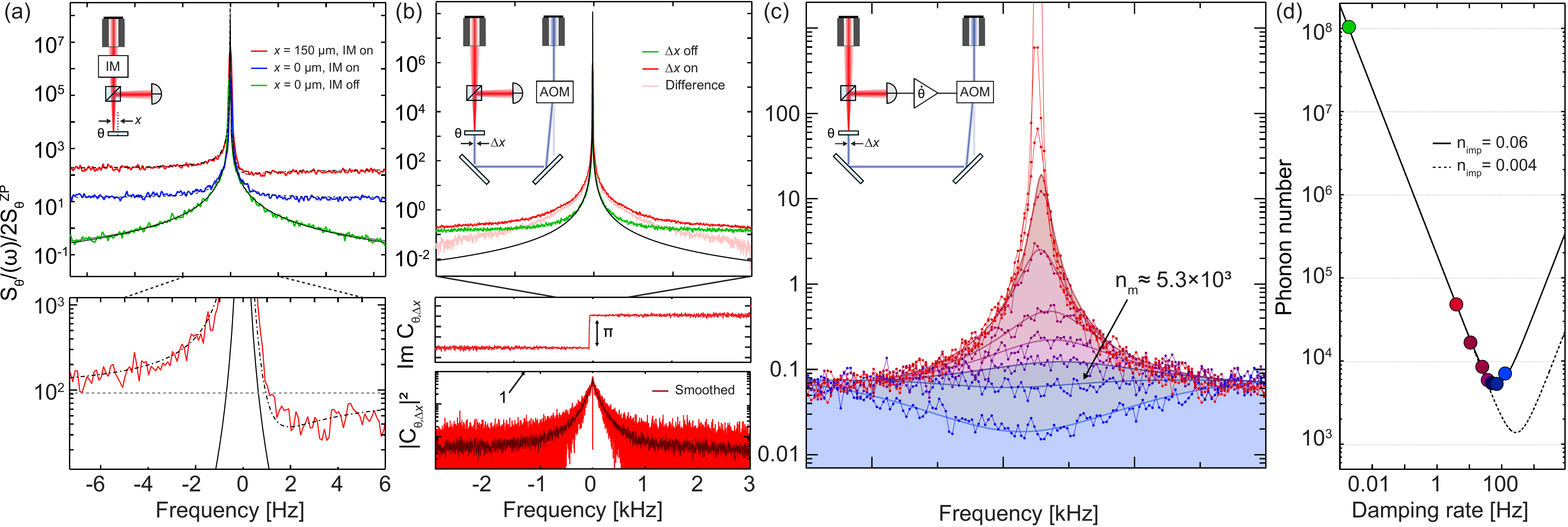}
		\caption{Radiation pressure backaction and control of a torsion oscillator. (a) Classical backaction and imprecision-backaction correlations produced by an intensity-modulated (IM) OL beam misaligned from the torsion axis. (b) Simulation of radiation pressure torque shot noise by position-modulating an auxiliary drive beam using an acousto-optic modulator (AOM).  Above: Thermal noise with (red) and without (green) auxiliary drive. Pink is the estimated backaction $S_\theta^\t{BA}$. Below: Coherence $C_{\theta\Delta x}$ between the OL measurement $\theta$ and drive beam displacement $\Delta x$ (see main text). (c) Cold damping the torsion mode by imprinting the OL signal onto the drive beam position with a $90^\circ$ phase shift near resonance. (d) Phonon number versus effective damping rate $\gamma_\t{eff}$ by fitting the noise spectra in (c) to the closed loop model $n_\t{m} = (\gamma_\t{eff}/\gamma_\t{m})(S_\theta(\omega_\t{m})+S_\theta^\t{imp})/(2 S_\theta^\t{ZP})$~ \cite{pluchar2020towards,wilson2015measurement}.}
		\label{fig:fig4}
		\vspace{-3mm}
	\end{figure*}

	In an effort to recover the ideal imprecision noise at large waist sizes, we investigated compensating the phase profile of the ribbon. 
	To this end, we adjusted the beam focus position $z$ to engineer a finite radius of curvature, $R(z) = z(1 + (z_R/z)^2)$ at the ribbon surface, where $z_R = \pi w_0^2/\lambda$ is the Rayleigh length. 
	As shown in Fig. \ref{fig:Fig3}b, we recorded $S_\theta^\t{imp}$ while varying the focus position with $w_0 = 60$ $\mu$m and $P = 100\,\mu\t{W}$, and found that the optimal position was indeed offset from the plane of the ribbon ($z = 0$).
	Overlaid is the same diffraction model as in Fig. \ref{fig:Fig3}b, indicating that for $w_0 \leq 64 \: \mu$m, we can fully compensate for the phase of the ribbon (for $w_0 > 64 \: \mu$m, the maximum Gaussian wavefront curvature $R(z_R) = 2 z_R$ is too small). 
	However, displacing the focus increases the spot size on the ribbon, thereby increasing the radiation pressure torque according to Eq. \ref{eq:StauQBA}. In our case, the offset is $z/z_R \approx 1$, so backaction is a factor of $w^2/w_0^2 = 2$ times larger than the minimum. 
	
	After optimizing the beam waist and focus position, we turned our attention to increasing optical power. Fig. \ref{fig:Fig3}c shows displacement spectra near mechanical resonance for several powers in the range $P = 0.01 - 10$ mW, calibrated to a thermal noise model (red line) \cite{pratt2023nanoscale}. In Fig. \ref{fig:Fig3}d, we plot $S_\theta^\t{imp}$ versus $P$ averaged over two spectral regions shaded in Fig. \ref{fig:fig2}b: relatively close to (green points) and far from (blue points) resonance, respectively.  Overlaid are models for the ideal imprecision and backaction of an optical lever (Eqs. \ref{eq:shotNoiseImpOL}-\ref{eq:StauQBA}) and fits to a model including a constant extraneous noise floor.
	For $P< 1\,\t{mW}$, we observe quantum noise scaling with an apparent total efficiency of $\eta = 48\%$, corresponding to a detector efficiency of $\eta_\t{d} = 75\%$.
	At higher powers, off-resonant thermal noise of nearby mechanical modes limits $S_\theta^\t{imp}$ in the region close to resonance to $2.7 \times 10^{-22}$ rad$^2$/Hz. In the region far from resonance, we continue to observe quantum noise down to $1.4\times 10^{-22}$ rad$^2$/Hz at $P = 10$ mW.  Scaling by $2S_\theta^\t{ZP}$ on the right axis yields (independent of the absolute value of $S_\theta^\t{ZP}$ for a thermal noise calibration) an minimum effective noise quanta of $n_\t{imp} = S_\theta^\t{imp}/(2S_\theta^\t{ZP}) = 0.004$, corresponding to an imprecision 18 dB below that the SQL ($n_\t{imp} = 1/4$) \cite{teufel2009nanomechanical,wilson2015measurement}. 
	
	We now turn our attention to radiation pressure backaction in an OL measurement.  To this end, as shown in Fig. \ref{fig:fig4}, we carried out a series of experiments using the auxialiary position-modulated laser to simulate a stochastic and coherent (dynamical) backaction torque.  
	
	We first emphasize that the OL can be made immune to classical backaction due to laser intensity noise---modulation of the HG$_{00}$ amplitude in Eq. \ref{eq:eRef}---when the probe beam is centered on the torsion axis. 
	We explored this by comparing OL measurements with different lateral beam positions.  As shown Fig. \ref{fig:fig4}a, intensity noise was increased by probing with an external cavity diode laser (ECDL) current modulated with white noise. When the beam was centered on the ribbon, we observed negligible backaction but increased imprecision, as we were unable to fully balance out the added intensity noise on the SPD. When the beam was displaced,	the total noise (physical motion and imprecision) increased and displayed an asymmetry about mechanical resonance.  This asymmetry is a signature of imprecision-backaction correlations mediated by the mechanical susceptibility, $S_{\theta, \tau} [\omega] \propto \t{Re}[\chi_\t{m}]$, a classical analog to ponderomotive squeezing \cite{marino2010classical}.  Fitting to a standard model \cite{sudhir2017quantum} (dashed black line) implies that classical intensity noise backaction overwhelms thermal noise $S_\tau^\t{BA,IM} [\omega] \approx 2.5  S_\theta^\t{th}$ and can be suppressed by at least an order of magnitude. 
	
	Quantum torque backaction arises due to vacuum  fluctuations of the HG$_{10}$ mode in Eq. \ref{eq:eRef}, physically manifesting as lateral beam fluctuations \cite{hao2024back,pluchar2024imaging}. 	To simulate this form of ``spatial" backaction \cite{pluchar2024imaging}, we applied white noise to the position-modulated drive beam until the motion it produced dominated the OL signal (Fig. \ref{fig:fig4}b). To confirm the backaction mechanism, we picked off a fraction of the drive beam and tracked its displacement $\Delta x$ on an auxiliary SPD. We then computed the cross spectrum $S_{\theta \Delta x} [\omega]$ with the OL signal $\theta$.  Fig. \ref{fig:fig4}b shows that the magnitude of the coherence $C_{\theta \Delta x}\equiv S_{\theta \Delta x} [\omega] /\sqrt{S_\theta [\omega] S_{\Delta x} [\omega]}$ approaches unity near the mechanical resonance, while the phase of the coherence (inset) exhibits a $\pi$ phase shift.  This behavior is consistent with mechanical motion dominated by radiation pressure torque noise \cite{purdy2013observation, pluchar2023thermal}.
	
	Combining quantum-limited measurement and coherent backaction (feedback) enables ground state preparation of a mechanical oscillator \cite{wilson2015measurement,rossi2018measurement}. To explore this possibility, as a final demonstration, we imprinted the OL measurement onto the drive beam position with an appropriate phase shift to realize cold damping \cite{poggio2007feedback,courty2001quantum}.  In the weak backaction limit, the phonon number of an oscillator cold damped at rate $\gamma_\t{eff}$ can be expressed as~\cite{wilson2015measurement}
	\begin{equation}
		n_\t{m} \approx \frac{\gamma_\t{m}}{\gamma_\t{eff}}n_\t{th}+\frac{\gamma_\t{eff}}{\gamma_\t{m}}n_\t{imp}\ge 2\sqrt{n_\t{th}n_\t{imp}}
	\end{equation}
	where $n_\t{th} = S_\theta^\t{th}/2S_\theta^\t{ZP} = k_B T/\hbar\omega_\t{m}$ is the thermal bath occupation.  Thus combining our $n_\t{imp} = 0.004$ OL measurement (Fig. \ref{fig:Fig3}d) and $n_\t{th} = 1.2\times 10^8$ torsion oscillator implies access to $n_\t{m} \approx 1.4\times 10^3$ from room temperature. Shown in Fig. \ref{fig:fig4}c-d is an experiment in which (for practical reasons related to the phase margin of our Red Pitaya controller \cite{neuhaus2023fpga}) we relax our imprecision to $n_\t{imp}\approx 0.06$ and demonstrate cold damping to $n_\t{m}\approx 5.3\times 10^3$.  Our data analysis procedure is described in~\cite{pluchar2020towards}.
	
	In summary, we have explored the quantum limits of OL measurement by probing the high $Q$ torsion mode of a Si$_3$N$_4$ nanoribbon. A key aim is to highlight the potential for torsional quantum optomechanics experiments.  Towards this end, we demonstrated a displacement imprecision 18 dB below that at the SQL, the working principle of radiation pressure shot noise in torque, and feedback cooling of a torsion oscillator from room temperature to $5.3\times 10^3$ phonons. In conjunction with cryogenics, the natural immunity of the OL to technical noise augurs well for future cavity-free quantum optomechanics experiments.  Indeed, at the time of this writing, we've become aware of a parallel study of Si$_3$N$_4$ nanoribbons with a ``mirrored" OL capable of imprecision at a level of $S_\theta^\t{imp}\sim 10^{-12}\,\t{rad}/\sqrt{\t{Hz}}$, by rejecting classical beam pointing noise \cite{shin2024laser}. Applied to optimized nanoribbons with $Q\approx 10^8$ \cite{pratt2023nanoscale} and reduced effective curvature (see Appendix) suggests that $n_\t{m}\sim 1$ may be accessible with this approach at cryogenic temperatures.  Scaling nanoribbons to the centimeter-scale \cite{cupertino2024centimeter} could also improve performance, owing to the favorable scaling law $S_\theta^\t{ZP}/S_\theta^\t{imp}\propto Q_0/w_\t{r} $ \cite{pratt2023nanoscale} afforded by torsional dissipation dilution and optical leverage.  Moreover, mass-loading Si$_3$N$_4$ nanoribbons has been shown not to diminish their torsional $Q$ \cite{pratt2023nanoscale}.  Thus, as emphasized by \cite{shin2024laser} and others \cite{agafonova2024laser,komori2020attonewton}, torsional optomechanics may be a promising route to milligram-scale quantum gravity experiments.
	\vspace{3mm}
	
	\emph{Note:} As mentioned, we recently became aware of a related
	independent study by Shin \emph{et. al.} \cite{shin2024laser}.
	
	\section*{Acknowledgements}
	The authors thank Wenhua He, Morgan Choi, Jon Pratt, Stephan Schlamminger, and Jack Manley for helpful discussions.  We also thank Atkin Hyatt and Mitul Dey Chowdhury for the photograph in Fig. \ref{fig:OL}. This work was supported by the National Science Foundation (NSF) through award nos. 2239735 and 2330310. CMP acknowledges support from the ARCS Foundation. ARA acknowledges support from a CNRS-UArizona iGlobes fellowship. The reactive ion etcher used for this study was funded by an NSF MRI grant, no. 1725571.
	
	\AtEndDocument{\vspace{-3mm}\section*{Appendix}
Here we provide details on various aspects of the theory, experiment and data analysis desribed in the main text. 


\vspace{-3mm}\section{Radiation pressure backaction torque}
In this section, we derive Eq. \ref{eq:StauQBA} and show that classical intensity noise produces zero net torque, 
using the semi-classical approach developed in Ref. \cite{pluchar2024imaging}. 

Consider a laser at normal incidence on a ribbon, centered along the axis of rotation (the $y$-axis) with photon intensity $I(x,y, t) = N(t) |u_{00}(x,y)|^2$,
where $u_{00}(x,y) = \sqrt{2/w(z)^2 \pi} e^{-(x^2 + y^2)/w(z)^2}$ is the Gaussian modeshape of the field with spot size $w(z)$. 
We assume that the length and width of the ribbon are much larger than $w(z)$, and therefore ignore any diffraction effects.

The net radiation pressure torque on the ribbon is given by
\begin{equation}
    \label{eq:tauRP}
    \tau_\t{RP}(t) = \frac{4 \pi \hbar}{\lambda} \int  I(x',y', t) x' dx' dy',
\end{equation}
where $x'$ is the distance from the axis of rotation to an incident photon. 
From Eq. \ref{eq:tauRP}, we observe the net torque is zero because the average photon position is zero (equal and opposite torques are applied to photons reflecting off either side of the ribbon), and thus the torsion motion is immune to classical radiation pressure intensity noise. 

To calculate the radiation pressure torque noise, we decompose the intensity into a mean and fluctuating component: $I(x,y,t) = I_0 + \delta I(x,y,t)$, where the fluctuations $\delta I(x,y,t)$ follow a Poisson distribution and the mean photon current is large. 
Assuming the photon arrivals of photons are uncorrelated in space, the single-sided cross spectral density can be expressed as 
\begin{equation}
    \label{eq:intensityCrossPSD}
    S_{\delta I,\delta I'}^\t{shot} = 2 I (x,y) \delta(x-x')\delta(y-y'),
\end{equation}
producing a fluctuating radiation pressure ($\delta P$)  cross spectral density of 
\begin{equation}
    S_{\delta P,\delta P'}^\t{shot} = \frac{16 \pi^2 \hbar^2}{\lambda^2} S_{\delta I,\delta I'}^\t{shot}.
\end{equation} 
The resulting backaction torque is found by integrating over all space
\begin{align*}
    \numberthis
    S_{\tau}^\t{BA} = \frac{16 \pi^2 \hbar^2}{\lambda^2} \iiiint (I(x,y) x dx dy) (I(x',y') x' dx' dy') = \frac{32 \pi^2 \hbar^2 N}{\lambda^2} \iint x^2 |u(x,y)|^2 dx dy,
\end{align*}
yielding Eq. \ref{eq:StauQBA}:
\begin{equation}
    \label{eq:OLBA}
    S_\tau^\t{BA} = \frac{8 N }{\theta_\t{D}^2}\frac{w^2}{w_0^2} \hbar^2  = w^2 \frac{2 h P}{c \lambda}.
\end{equation}

\vspace{-3mm}\section{Optical phase profile of the nanoribbon}

The maximum sensitivity of our OL measurements is constrained by the anomalous phase profile of the nanoribbon. 
In this section, we discuss the details of the white light interferometer measurements we made of the phase profile of the nanoribbon used in the experiment, as well as measurements of similar devices.

For the measurement in Fig. \ref{fig:fig2}b, we fit the data with the equation 
\begin{equation}
    y = A_0 + A_l x + A_p x^2,
\end{equation}
and find the dominant contribution is due to the quadratic term, indicating that for larger waist sizes, the nanoribbon surface acts like a parabolic mirror.  
This focuses the reflected light to a small waist size, increasing the divergence angle and resulting in larger imprecision noise (Eq. \ref{eq:shotNoiseImpOL}). 
This has a smaller effect on beams with smaller spot sizes, as the surface appears locally flat. 

For other ribbon geometries, we find the radius of curvature of the phase can be reduced. 
In addition to the nanoribbon used in the experiment, we also measured the anomalous curvature of several other nanoribbons with a white light interferometer, shown in Fig. \ref{fig:ribbonCurvatureComp}. 
Two features are notable: (1) For smaller ribbon widths, the measured radius of curvature decreases, which we predict will reduce the OL sensitivity (see the next section); and (2) for nanoribbons with fillet geometries numerically designed to maximize the torsion mode $Q$ (not studied here or in \cite{pratt2023nanoscale}), but pictured in Fig. \ref{fig:OL}, the measured phase profile is closer to planar.  The latter may help realize lower imprecision OL measurements with larger waist sizes. 


\vspace{-3mm}\section{Fraunhofer diffraction model}

Here, we examine the consequences of the nonplanar phase profile of the nanoribbon on the OL sensitivity by utilizing a Fraunhofer diffraction model, shown in Fig. \ref{fig:Fig3}b. 

A parabolic reflector with width $w_\t{r}$ and length $L \gg w_\t{r}$ is placed at the focus of a Gaussian beam.
An SPD is located at a distance $L_\t{OL} \gg z_0$ away from the reflector, where the Fraunhofer approximation is valid. 
The field at the detector can be expressed as
\begin{align}
    \label{eq:Eff}
    E_\t{r}(x, y, z = L_\t{OL}) & \propto \int^\infty_{-\infty}  dy' \int^{w_\t{r}/2}_{-w_\t{r}/2} dx'  e^{-(x'^2 + y'^2)/w_0^2 + e^{i 2 A_p x'^2} - i \frac{2 \pi}{L_\t{OL} \lambda}(x x' + y y')}.
\end{align}
We then numerically compute the output of the SPD, given by 
\begin{equation}
\label{eq:sensitivityIntegral}
\Delta P(x) = \int_{-\infty}^{x} |E_\t{r}(x',y')|^2 dx'dy' - \int_{x}^{\infty} |E_\t{r}(x',y')|^2 dx'dy',
\end{equation}
which we use to calculate the shot-noise-limited imprecision noise from \cite{pratt2023nanoscale}
\begin{equation}
    \label{eq:SthetatImpApp}
    S_\theta^\t{imp} = \left(\frac{\partial \Delta P}{\partial x}\frac{\partial x}{\partial \theta}\right)^{-2} S_P^\t{shot},
\end{equation}
where $S_P^\t{shot} = 4\pi \hbar c P / \lambda$ is the laser power shot noise spectrum.

\vspace{-3mm}\section{Position-dependent phase compensation with a Gaussian beam}

In order to reduce the disparity between the ideal and measured imprecision in Fig. \ref{fig:Fig3}b, we used the parabolic wavefront of the Gaussian beam to compensate for the observed parabolic phase of the nanoribbon, shown in Fig. \ref{fig:Fig3}c.
Using the same model discussed in the previous section, we now allow the beam focus to be displaced from the focus by a distance $z'$, and correspondingly include the Gaussian beam wavefront radius of curvature into Eq. \ref{eq:Eff}.
The field at the SPD is now given by 
\begin{equation}
    \label{eq:Eff}
   E_\t{r}(x, y, z = L_\t{OL}) \propto \int^\infty_{-\infty}  dy' \int^{w_\t{r}/2}_{-w_\t{r}/2} dx'  e^{-(x'^2 + y'^2)/w(z')^2 + e^{i 2 A_p x'^2} + \pi(x'^2 + y'^2)/(\lambda R(z)) - i \frac{2 \pi}{L_\t{OL} \lambda}(x x' + y y') },
\end{equation}
and we compute the optical lever (OL) imprecision using Eqs. \ref{eq:sensitivityIntegral}-\ref{eq:SthetatImpApp}. 
For the measurements in Fig. \ref{fig:Fig3}c, $w_0 = 60 \: \mu$m, indicating that near $z' = -15$ mm, the parabolic curvature of the beam and ribbon cancel, yielding planar wavefronts, restoring the ideal sensitivity.

\begin{figure}
    \centering
    \includegraphics[width=0.5\linewidth]{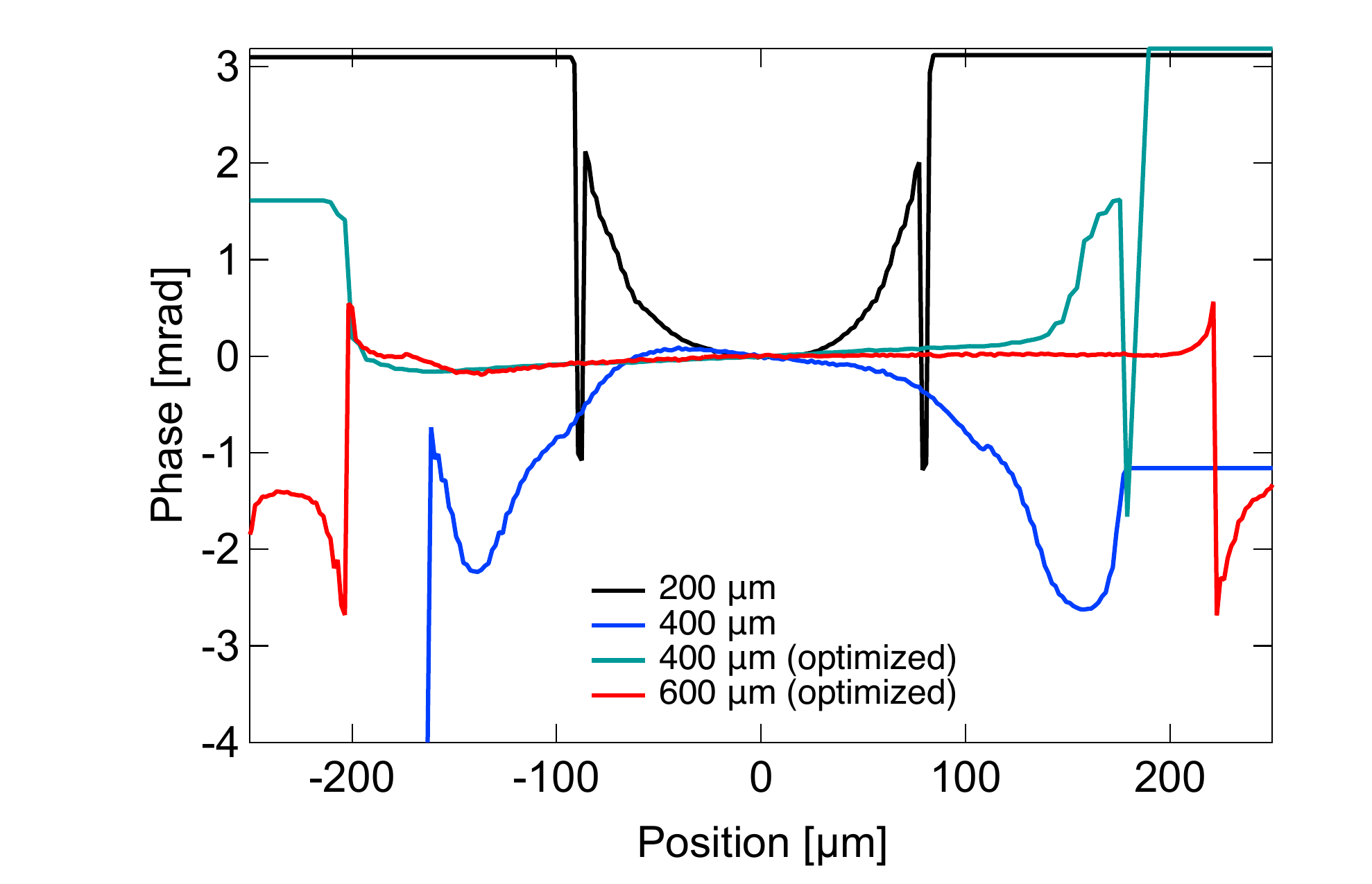}
    \caption{Phase profile of nanoribbons with different geometries.}
    \label{fig:ribbonCurvatureComp}
    \vspace{-3mm}
\end{figure}

\section{Calibration and Photothermal heating}

For the data presented in Figs. \ref{fig:Fig3}-\ref{fig:fig4}, we calibrate the measurements by bootstrapping to a thermal noise model of the torsion mode, $S_\theta^\t{th} = S_\tau^\t{th}|\chi_\t{m}|^2$ \cite{pratt2023nanoscale}. 
Photothermal heating can lead to systematic errors in calibration, as well as an increase in the starting thermal occupation number. 
To determine if the torsion mode is subject to photothermal heating at larger incident optical powers, we perform a second, independent, calibration by directly measuring the lateral displacement sensitivity $\partial V / \partial \Delta x$ by sweeping the position of the split photodetector by known amounts, shown in Fig. \ref{fig:calibrationComp}a.
We then calculate the angular displacement sensitivity via \cite{pratt2023nanoscale}
\begin{equation}
    \frac{\partial V}{\partial \theta} = 2 L_\t{OL} \frac{\partial V}{\partial \Delta x}. 
\end{equation}
A comparison between the two calibration methods (Fig. \ref{fig:calibrationComp}b) at large optical powers suggests the photothermal heating is below the uncertainty in our calibrations. 

\begin{figure}
    \centering
    \includegraphics[width=1\linewidth]{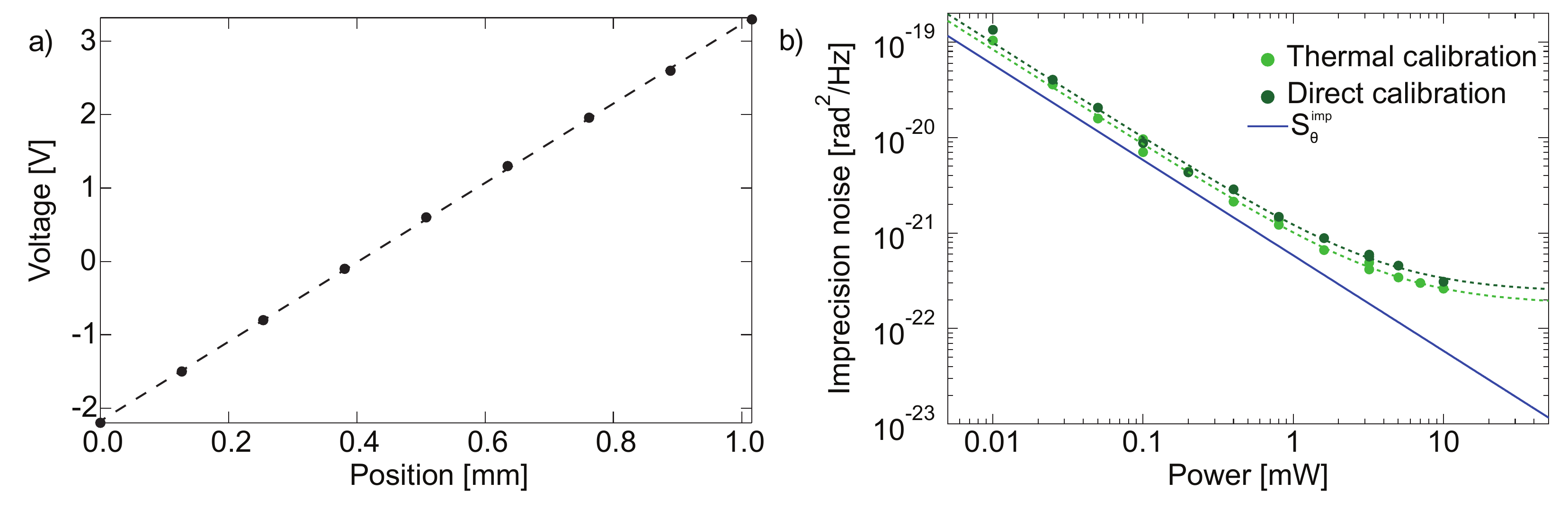}
    \caption{a) Measurement of $\partial V / \partial \Delta x$ for $P = 25 \: \mu$m. b) Comparison of the calibration methods for the data shown in Fig. \ref{fig:Fig3}a and d. Dashed lines are fits to the data.}
    \label{fig:calibrationComp}
\end{figure}

\section{Classical imprecision-backaction correlations}

In this section, we discuss the details of the model in Fig. \ref{fig:fig4}a. 
In the regime of strong radiation pressure noise, the components of Eq. \ref{eq:eq2} are no longer uncorrelated, as the fluctuations of the laser (in this case, in the form of classical intensity noise) contribute both imprecision and backaction noise to the measurement. 
As such, we include the cross-spectral density between the imprecision and backaction noise $S_{\theta, \tau} [\omega] = C \t{Re}[\chi_\t{m}]$, where $C$ is a factor that accounts for the measurement strength and the measured quadrature angle (which is slightly offset from the phase quadrature), to Eq. \ref{eq:eq2} \cite{sudhir2017quantum}:
\begin{equation}
    S_\theta = S_\theta^\t{imp} + (S_\tau^\t{BA, IM} + S_\tau^\t{th})|\chi_\t{m}|^2 + S_{\theta, \tau} [\omega]. 
\end{equation}
In fitting the data, $S_\theta^\t{imp}, S_\tau^\t{BA, IM},$ and $C$ are free parameters, as $S_\tau^\t{th}$ is independently fixed by calibrating the data to a fixed radiation pressure tone applied by the auxiliary laser.

}
	
	\bibliography{ref}
\end{document}